\documentclass[aps,prd,amssymb,nofootinbib,twocolumn,epsf,floatfix,showpacs]
{revtex4}
\usepackage[usenames]{color} 
\usepackage{ulem} 
\usepackage{graphicx}
\usepackage{amssymb}
\usepackage{amsmath}
\usepackage{epstopdf}

\begin{document}


\title{Causal Representations of Neutron-Star
  Equations of State}

\author{Lee Lindblom}

\affiliation{Center for Astrophysics and Space Sciences, University 
of California at San Diego, La Jolla, CA 92093, USA}

\date{\today}
 
\begin{abstract}
  Parameterized representations of the equation of state play an
  important role in efforts to measure the properties of the matter in
  the cores of neutron stars using astronomical observations.  New
  representations are presented here that are capable of representing
  any equation of state to any desired level of accuracy, while
  automatically imposing causality and thermodynamic stability
  constraints.  Numerical tests are presented that measure how
  accurately and efficiently these new parameterizations represent a
  collection of causal nuclear-theory model equations of state.
\end{abstract}


\maketitle


\section{Introduction}
\label{s:introduction}
The equation of state describes the relationship between the total
energy density $\epsilon$, the pressure $p$, the temperature $T$, and
any other variables that determine the thermodynamic state of the
material in a fluid.  In newly formed neutron stars the extremely hot
matter is expected to evolve quickly to its lowest energy state
through high energy particle interactions; and the temperature of this
material is expected to drop well below the Fermi temperature on a
fairly short timescale through neutrino and photon emissions.  Under
these conditions the equation of state rapidly becomes almost
barotropic, i.e. the energy density depends essentially only on the
pressure of the material: $\epsilon=\epsilon(p)$.

The density of most of the material in a neutron star is larger than
that of the matter in an atomic nucleus.  Such densities are well
beyond those accessible to laboratory experiments.  Heavy ion
scattering experiments (such as those carried out at RHIC and LHC)
provide important information about the interactions between the
various particles that make up neutron-star
matter~\cite{CBMPhysicsBook}.  However those experiments can not
duplicate the low temperature (i.e. well below the Fermi temperature)
but very high density equilibrium conditions that exist in neutron
stars.

\begin{figure}[!tb]
\centerline{\includegraphics[width=3in]{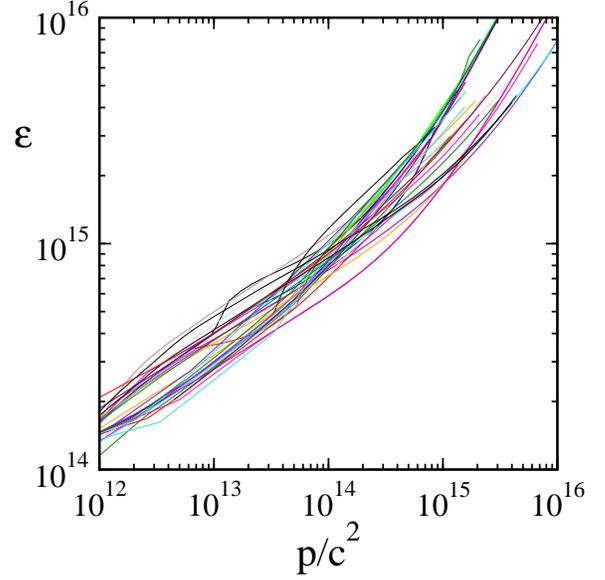}}
\caption{\label{f:RealisticEOS} Curves illustrate 34 nuclear-theory
  models of the neutron-star equation of state,
  $\epsilon=\epsilon(p)$.  The density $\epsilon$ and pressure $p$
  (divided by the speed of light squared) are displayed here in cgs
  units, g/cm${}^3$.}
\end{figure}

In contrast to the dearth of directly relevant experimental data, a
great deal of theoretical work has been done using the available data
to construct models of neutron-star matter.  Unfortunately there is no
consensus at present on {\it the} correct model.  This is not
surprising, given the significant differences between neutron-star
matter and the better understood matter contained in atomic nuclei.
In particular, complicated many body interactions are likely to play
an extremely important role in determining the properties of the cold
very dense neutron-star matter. Figure~\ref{f:RealisticEOS}
illustrates 34 published equations of state, $\epsilon=\epsilon(p)$,
for neutron-star matter based on a variety of nuclear theory
models.\footnote{The 34 equations of state shown in
  Fig.~\ref{f:RealisticEOS} are those used by Read, et
  al.~\cite{Read:2008iy} in their study of the piecewise-polytropic
  representations of neutron-star equations of state.  The abbreviated
  names of these equations of state are: PAL6, SLy, APR1, APR2, APR3,
  APR4, FPS, WFF1, WFF2, WFF3, BBB2, BPAL12, ENG, MPA1, MS1, MS2,
  MS1b, PS, GS1, GS2, BGN1H1, GNH3, H1, H2, H3, H4, H5, H6, H7, PCL2,
  ALF1, ALF2, ALF3.  ALF4.}  This figure illustrates the lack of
consensus among the various models: at a given density, $\epsilon$,
the associated value of the pressure varies by almost an order of
magnitude.

Given the lack of direct laboratory access to neutron-star matter and
the lack of consensus theoretical modeling for this material,
observations of neutron stars may turn out to be the most promising
way to obtain quantitative information about the equation of state of
neutron-star matter.  It is well known that the equation of state
together with the gravitational field equations determine the
structures and hence all the observable macroscopic properties of
neutron stars~\cite{Oppenheimer1939}.  Conversely, it is also known
that a complete knowledge of an appropriate set of macroscopic
properties (e.g.  masses and radii) together with the gravitational
field equations uniquely determines the equation of
state~\cite{Lindblom1992, Lindblom2014a}.  Studies of neutron-star
models have shown that their macroscopic properties, like masses and
radii, vary widely even within the current ``realistic'' class of
equations of state~\cite{Read:2008iy, ArnettBowers1977}.  Observations
of the macroscopic properties of these stars would therefore provide
significant constraints on the equation of state.  Unfortunately the
needed observations are quite difficult to make.  Masses of several
dozen neutron stars have now been measured fairly accurately (see
e.g. Ref.~\cite{Lattimer2007}).  These observations alone rule out
large classes of very ``soft'' equation of state models that fail to
support sufficiently massive neutron stars.  A number of radius
measurements of neutron stars have also been made~\cite{Ozel2009a,
  Steiner2010, Guver2010a, Guver2010, Steiner2012, Galloway2012,
  Guver2012a, Guver2012b, Zamfir2012, Ozel2012, Guver2013,
  Lattimer2014, Nattila2017, Steiner2018, Shaw2018}.  At present these
radius measurements are not nearly as accurate as the best mass
observations, but hopefully the quantity and quality of these
observations will continue to improve over time.

Good simultaneous measurements of both the mass and the radius (or
some other macroscopic observable like the tidal deformability or the
moment of inertia) are needed to determine the equation of state
quantitatively from observations of neutron stars.  Given a suitable
set of observations, $M_i$ and $R_i$ (for example), the equation of
state can be determined in the following way.  First, a parametric
representation of possible equations of state,
$\epsilon=\epsilon(p,\lambda_k)$, is chosen.  Next, stellar models are
constructed using this parametric equation of state, and the
associated observables, $M(\lambda_k)$ and $R(\lambda_k)$, are
evaluated for these models.  Finally, the parameters $\lambda_k$ are
adjusted to optimally match the model observables, $M(\lambda_k)$ and
$R(\lambda_k)$, with the actual observations, $M_i$ and $R_i$.  The
resulting equation of state, $\epsilon=\epsilon(p,\lambda_k)$, with
this optimal choice of $\lambda_k$ is an approximate representation of
the physical neutron-star equation of state.

The approach outlined above for determining the equation of state has
been tested using mock observational data constructed from the
nuclear-theory model equations of state~\cite{Lindblom1992,
  Lindblom2012, Lindblom2014, Lindblom2014a, Abdelsalhin2018}.  These
tests show that accurate measurements of observables from a small
number (just two or three) neutron stars can determine the high
density part of the equation of state in the cores of neutron stars at
the few percent accuracy level.  Several groups have used
qualitatively similar methods to estimate the actual neutron-star
equation of state with the currently available (fairly inaccurate)
observational data~\cite{Ozel2010, Steiner2010, Steiner2012, Ozel2016,
  Ozel2016a, Raithel2016a, Raithel2016b, Nattila2017}.

This paper studies one important technical aspect of the method
outlined above for determining the neutron-star equation of state: how
to choose an appropriate parametric representation $\epsilon=
\epsilon(p, \lambda_k)$.  Several basic principles inform this choice:
\begin{enumerate}
\item Any equation of state should be representable by some
  $\epsilon=\epsilon(p,\lambda_k)$ to any desired level of accuracy.
\item Each $\epsilon=\epsilon(p,\lambda_k)$ should satisfy the basic
  laws of physics, like thermodynamic stability and causality.
\item The representation should be efficient.
\end{enumerate}
Since the neutron-star equation of state is not well understood at
this time, it would not be appropriate to restrict its possible form
with a non-universal representation.  Principle \#1 ensures that this
does not happen.  Whatever the neutron-star equation of state turns
out to be, we can be quite confident that it will satisfy the basic
laws of physics like thermodynamic stability and causality.  Principle
\#2 requires that each parametric representation satisfies these
conditions.  And finally Principle \#3 states the practical need for
efficient parametric representations, i.e. those capable of achieving
good accuracy with the fewest number of parameters possible.  Fixing
two parameters experimentally to achieve a certain level of accuracy
with an efficient parameterization, for example, is much better than
needing to fix four parameters to achieve the same level of accuracy
using a less efficient parameterization.

Previous efforts to construct suitable representations of the
neutron-star equation of state satisfy some, but not all of these
principles.  The widely used piecewise-polytropic
representations~\cite{Vuille1995, VuilleIpser1999, Read:2008iy} satisfy
Principle \#1, and the thermodynamic stability part of Principle \#2,
but they are not particularly efficient.  The spectral
representations~\cite{Lindblom2010} are more efficient, and they also
satisfy Principle \#1, and the thermodynamic stability part of
Principle \#2.  None of the published representations enforce
causality.

The main goal of this paper is to introduce and test new
representations that satisfy all of the principles described above,
including in particular the causality condition.
Section~\ref{s:FaithfulEOS} reviews the derivations of representations
that satisfy Principle \#1, and the thermodynamic stability part of
Principle \#2.  Section~\ref{s:CausalEOS} presents generalizations of
these methods that find representations enforcing the simple causality
condition appropriate for true barotropic equations of state.
Section~\ref{s:EnthalpyBasedForms} expands these discussions to
include enthalpy-based representations that are the analogs of the
pressure-based representations discussed in Secs.~\ref{s:FaithfulEOS}
and \ref{s:CausalEOS}.  Section~\ref{s:RealisticEOS} tests the
accuracy and efficiency of the various representations considered here
(including the previously published forms) by constructing and
measuring the accuracy of representations of 27 causal nuclear-theory
model equations of state, the causal subset of those studied by Read,
et al.~\cite{Read:2008iy}.  And finally Sec.~\ref{s:Discussion} gives
a brief summary and discussion of the most interesting and important
results.


\section{Thermodynamically Stable Representations}
\label{s:FaithfulEOS}

Thermodynamic stability requires that the energy density $\epsilon$
must be a non-negative and monotonically increasing function of the
pressure: $\epsilon(p_1)\ge\epsilon(p_0)\geq 0$ for $p_1\ge p_0\ge0$.
For sufficiently differentiable equations of state, this is equivalent
to the condition that $d\epsilon/dp\geq 0$.  Every (sufficiently
differentiable) equation of state $\epsilon=\epsilon(p)$ determines
a unique adiabatic index $\Gamma(p)$,
\begin{eqnarray}
  \Gamma(p)=\frac{\epsilon(p)+p}{p}\left[\frac{d\epsilon(p)}{dp}\right]^{-1},
\end{eqnarray}
and thermodynamic stability therefore requires $\Gamma(p)$ to be
non-negative: $\Gamma(p)\geq 0$.  Conversely every non-negative
function $\Gamma(p)$ determines a unique (up to a constant of
integration) equation of state $\epsilon=\epsilon(p)$ that satisfies
the thermodynamic stability condition.  This equation of state is the
solution of the ordinary differential equation
\begin{eqnarray}
  \frac{d\epsilon(p)}{dp} = \frac{\epsilon(p)+p}{p\,\Gamma(p)}.
  \label{e:Gammaode}
\end{eqnarray}
The problem of constructing representations of the equation of state,
$\epsilon(p)$, is therefore equivalent to the problem of
constructing representations of $\Gamma(p)$.

Physically acceptable representations of the adiabatic index
$\Gamma(p)$ are easier to construct than representations of the
equation of state $\epsilon(p)$ itself.  Acceptable $\epsilon(p)$ are
more restrictive: they must be non-negative and monotonically
increasing, while the $\Gamma(p)$ need only be non-negative. Two
different approaches have been used to construct appropriate
representations of $\Gamma(p)$.  The first constructs
piecewise-analytical approximations for $\Gamma(p)$, which can be made
arbitrarily accurate using a sufficiently large number of parameters.
The piecewise-polytropic representations~\cite{Vuille1995,
  VuilleIpser1999, Read:2008iy} are an example of this approach.  The
other approach constructs a spectral representation of $\log
\Gamma(p)$, which can also be made arbitrarily accurate by including a
sufficient number of terms in the expansion.  An example of this
approach~\cite{Lindblom2010} is a representation of $\log\Gamma(p)$ as
polynomials in $\log p$.

The basic idea of the piecewise-analytical representations is to use
simple analytical functions $\Gamma(p,\gamma_k)$ to represent
$\Gamma(p)$ on each of $n$ subdomains, $p_k\leq p < p_{k+1}$, of the
pressure domain of interest, $p_0\leq p \leq p_\mathrm{max}$. The
piecewise-polytropic representations are a particularly simple
example.  In this case the functions $\Gamma(p,\gamma_k)$ are taken
simply to be non-negative constants in each interval:
$\Gamma(p,\gamma_k) =\Gamma(p_k) =\gamma_k\geq 0$ for $p_{k} \leq p <
p_{k+1}$.  Any $\Gamma(p)$ can clearly be approximated to any desired
degree of accuracy in this way by using a sufficiently large number of
pressure subdomains.

Once a particular representation $\Gamma=\Gamma(p,\gamma_k)$ is
chosen, it is straightforward to recover the underlying equation of
state $\epsilon=\epsilon(p,\gamma_k)$ by solving
Eq.~(\ref{e:Gammaode}).  As shown in Ref.~\cite{Lindblom2010} the
solution to this equation can be reduced to quadratures:
\begin{eqnarray}
\epsilon(p)= \frac{\epsilon_0}{\mu(p)}+
\frac{1}{\mu(p)}\int_{p_0}^p\frac{\mu(p')}{\Gamma(p')}dp',
\label{e:EpsilonGamma} 
\end{eqnarray}
where $\mu(p)$ is defined as
\begin{eqnarray}
  \mu(p)=\exp\left[-\int_{p_0}^p \frac{dp'}{p'\,\Gamma(p')}\right],
  \label{e:MuGamma}
\end{eqnarray}
and where $\epsilon_0=\epsilon(p_0)\geq 0$ is the constant of
integration needed to fix the solution.  For the simple
piecewise-polytropic representation these integrals can be done
analyticly, and the equation of state $\epsilon(p,\gamma_k)$ has the
very simple analytical form,
\begin{eqnarray}
  \epsilon(p,\gamma_k)=\left(\epsilon_k - \frac{p_k}{\gamma_k -1}\right)
  \left(\frac{p}{p_k}\right)^{1/{\gamma_k}}
  +\frac{p}{{\gamma_k} -1},
\end{eqnarray} 
in the domain where $p_k\leq p \leq p_{k+1}$.  The constants $\epsilon_k$
are determined by the recursion relations
\begin{eqnarray}
  \epsilon_{k+1}=\left(\epsilon_k - \frac{p_k}{{\gamma_k} -1}\right)
  \left(\frac{p_{k+1}}{p_k}\right)^{1/{\gamma_k}}
  +\frac{p_{k+1}}{{\gamma_k} -1}
\end{eqnarray}
that enforce continuity of $\epsilon(p,\gamma_k)$ at the
subdomain boundaries.

The basic idea of spectral representations is to express functions as
linear combinations of a complete set of basis functions,
e.g. $\Gamma(p)=\sum_k \gamma_k\Phi_k(p)$.  Physically acceptable
$\Gamma(p)$ must be non-negative, so unfortunately straightforward
spectral expansions of $\Gamma(p)$ are not useful. Typical basis
functions, like the trigonometric functions or the orthogonal
polynomials, are oscillatory, therefore allowing $\Gamma(p)$ to become
negative for many choices of the $\gamma_k$.  This problem can be
avoided, however, by using the spectral expansion to determine
$\log\Gamma(p)$:
\begin{eqnarray}
  \Gamma(p,\gamma_k) = \exp\left[\sum_k\gamma_k \Phi_k(p)\right].
\end{eqnarray}
Any complete set of basis functions $\Phi_k(p)$ can be used. The
functions $\Gamma(p,\gamma_k)$ are automatically non-negative for any
choice of $\gamma_k$; and these representations can be made
arbitrarily accurate simply by including enough terms in the
expansion.  As in the piecewise-analytical case, the equation of state
$\epsilon=\epsilon(p,\gamma_k)$ is recovered using
Eqs.~(\ref{e:EpsilonGamma}) and (\ref{e:MuGamma}).  While the
integrals in these equations can not in general be done analyticly,
they can be done numerically very efficiently.  The simple choice of
basis functions $\Phi_k(p)=[\log(p/p_0)]^k$ was tested and used in
Ref.~\cite{Lindblom2010}.

\begin{figure}[h]
\centerline{\includegraphics[width=3in]{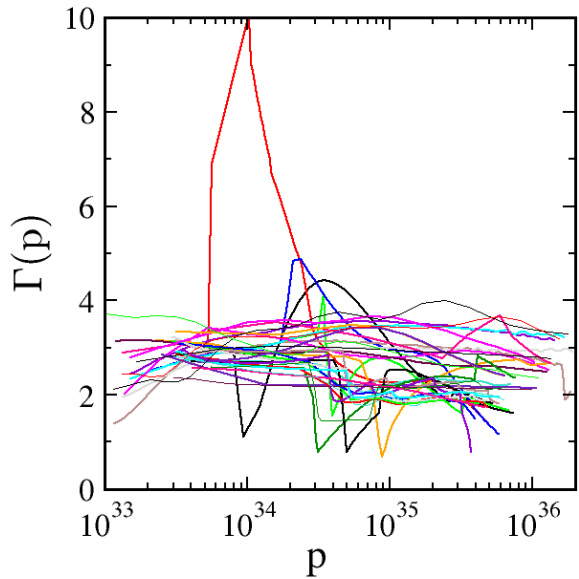}}
\caption{\label{f:AdiabaticIndex} Curves illustrate the adiabatic
  index $\Gamma(p)$ for the 27 causal nuclear-theory model equations
  of state used to test the accuracy and efficiency of various
  parametric representations.}
\end{figure}

Figure~\ref{f:AdiabaticIndex} illustrates $\Gamma(p)$ for the 27
causal nuclear-theory model neutron-star equations of state used as
test examples in this paper.  These equations of state include all
those used by Read, et al.~\cite{Read:2008iy} in their study of the
piecewise-polytropic representations, except the 7 which violate the
causality condition $dp/d\epsilon\leq c^2$ for some range of
densities: APR2, APR3, APR4, WFF1, WFF2, ENG, and ALF1.  The
particular equations of state have not been identified with their
corresponding curves in Fig.~\ref{f:AdiabaticIndex} (nor those in
Figs.~\ref{f:SoundSpeeds}--\ref{f:h_causal_spec_chisqr_nlambda}).
This omission was motivated in part by practical considerations: there
are simply too may closely spaced and overlapping curves to allow them
to be labeled clearly.  Instead, the data presented in these figures
is intended to give an overall impression in
Figs.~\ref{f:AdiabaticIndex}--\ref{f:SoundSpeedsh} of the range of
variability in the nuclear-theory model equations of state, and an
overall impression in
Figs.~\ref{f:p_causal_simp_chisqr_nlambda}--\ref{f:h_causal_spec_chisqr_nlambda}
of the range of accuracies of the various representations studied
here.  The detailed discussion of how efficiently the
piecewise-analytical (PB-PA) and the spectral (PB-S)
representations\footnote{The notation PB-PA is used to denote the
  pressure-based piecewise-analytical representation, while PB-S
  denotes the pressure-based spectral representation.} work for these
various equations of state is delayed to Section~\ref{s:RealisticEOS}.
However to summarize, those tests reveal that the spectral expansions
are more efficient, but both approaches work surprising well given the
complexity of the functions $\Gamma(p)$ seen in
Fig.~\ref{f:AdiabaticIndex}.
%


\section{Causal Representations}
\label{s:CausalEOS}

Sound waves propagate at the ``characteristic'' speeds of the
hyperbolic equations that describe the evolution of a relativistic
fluid.  For a fluid having a true barotropic equation of state
(i.e. one that is barotropic even for non-equilibrium fluid states),
these characteristic speeds are related to the equation of state by
the simple expression
\begin{eqnarray}
v^2(p)=\left[\frac{d\epsilon(p)}{dp}\right]^{-1}.
\end{eqnarray}
Causality, therefore, requires the equation of state for such fluids to
satisfy the requirement $d\epsilon(p)/dp \geq c^{-2}$, where $c$ is the
speed of light.

Neutron-star matter is a complex mixture of many species of particles.
There is a reasonably compelling argument that the equilibrium state
of this material has a barotropic equation of state (see Sec. I).
However the argument that the equation of state remains barotropic
even for the non-equilibrium perturbations that propagate as sound
waves is much less compelling.\footnote{The abundances of the various
  particles that make up neutron-star matter are determined in part by
  the conditions for beta equilibrium.  These abundances will not have
  time to adjust in high frequency perturbations to their equilibrium
  values via these relatively slow weak interactions.  The sound speed
  may therefore be somewhat different from that of a pure barotrope.}
The characteristic speeds of such complex mixtures will be determined
by the detailed micro-physics that governs the dynamics of the
mixture.  Given the lack of a consensus nuclear-theory model for
neutron-star matter, it is not surprising that the characteristic
speeds and hence the causal properties of this material are not well
understood at this time.  Given this uncertainty, this paper simply
adopts the (widely used but possibly irrelevant) causality condition
for true barotropes:
\begin{equation}
  \frac{d\epsilon(p)}{dp} =\frac{1}{v^2(p)}\geq \frac{1}{c^{2}}.
  \label{e:causality}
  \end{equation}
\hyphenation{baro-tropic}

The goal in this section is to construct parametric representations of
the equation of state that automatically enforce all of the principles
outlined in Sec.~\ref{s:introduction}, including causality.
Thermodynamic stability is equivalent to the condition that the sound
speed squared, $v^2(p)$, is non-negative.  This condition, together
with the causality condition given in Eq.~(\ref{e:causality}), implies
that any physical equation of state must have sound speeds that
satisfy, $0\leq v^2(p)\leq c^2$.  It is convenient to impose both the
causality and the thermodynamic stability conditions as a single
inequality on the function $\Upsilon(p)$:
\begin{eqnarray}
 \Upsilon(p)=\frac{c^2-v^2(p)}{v^2(p)}\geq 0.
\end{eqnarray}

The function $\Upsilon(p)$ plays much the same role in causal
representations as the adiabatic index $\Gamma(p)$ played in the
discussion of the thermodynamically stable representations of the
equation of state in Sec.~\ref{s:FaithfulEOS}.  Like the adiabatic
index, $\Upsilon(p)$ must be non-negative, $\Upsilon(p)\geq 0$. Like
the adiabatic index, every (sufficiently differentiable) equation of
state $\epsilon=\epsilon(p)$ determines a unique $\Upsilon(p)$. And
like the adiabatic index, every (sufficiently smooth) non-negative
function $\Upsilon(p)$ determines a unique (up to a constant of
integration) equation of state $\epsilon=\epsilon(p)$.  This equation
of state is the solution of the ordinary differential equation
\begin{eqnarray}
  \frac{d\epsilon(p)}{dp}=\frac{1}{c^{2}} + \frac{\Upsilon(p)}{c^{2}}.
  \label{e:EpsilonP}
\end{eqnarray}
The solution to this equation is given by the simple quadrature:
\begin{eqnarray}
  \epsilon(p) = \epsilon_0 + \frac{p-p_0}{c^2}+\frac{1}{c^{2}}\int_{p_0}^p
  \Upsilon(p')dp',
  \label{e:EpsilonVelocity}
\end{eqnarray}
where $\epsilon_0$ is the constant of integration
$\epsilon_0=\epsilon(p_0)\geq 0$ needed to fix the solution.

This one-to-one correspondence between non-negative $\Upsilon(p)$ and
equations of state satisfying the causality and thermodynamic
stability conditions provides a natural route to constructing causal
representations of these equations of state.  Every non-negative
$\Upsilon(p)$ generates a causal representation.  As shown in
Sec.~\ref{s:FaithfulEOS}, such functions are easy to approximate at
any level of accuracy using a variety of methods, including
piecewise-analytical approximations and spectral approximations.

\begin{figure}[!t]
\centerline{\includegraphics[width=3in]{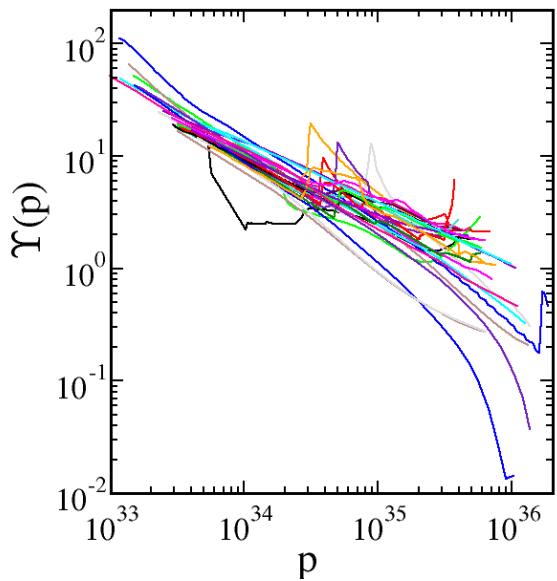}}
\caption{\label{f:SoundSpeeds} Curves illustrate the sound speed
  function $\Upsilon(p)=[c^2-v^2(p)]/v^2(p)$ for 27 nuclear-theory
  model equations of state that satisfy the causality condition,
  $\Upsilon(p)\geq 0$, in the pressure domain of interest to
  neutron-star interiors.}
\end{figure}

Consider first a piecewise-analytical representation constructed by
subdividing the pressure domain $[p_0,p_\mathrm{max}]$ into $n$
subdomains with $p_0<p_1<...<p_{n-1}<p_\mathrm{max}$.
Figure~\ref{f:SoundSpeeds} illustrates the functions $\Upsilon(p)$ for
the 27 causal nuclear-theory model equations of state in the pressure
domain of interest for neutron-star interiors.  This figure shows that
$\log\Upsilon(p)$ is roughly proportional to $\log p$ for these
equations of state.  A reasonably good analytical representation of
$\Upsilon(p)$ in each pressure subdomain is therefore given by
\begin{equation}
 \Upsilon(p,\upsilon_k)=
 \Upsilon_k\left(\frac{p}{p_k}\right)^{\upsilon_{k+1}},
\end{equation}
where the adjustable parameters $\Upsilon_k$ and $\upsilon_{k+1}$
determine its properties in each subdomain.  Given this representation
of $\Upsilon(p)$, the integral in Eq.~(\ref{e:EpsilonVelocity}) is
easy to perform, and the resulting $\epsilon(p,\upsilon_k)$ is given
by
\begin{eqnarray}
  \epsilon(p,\upsilon_k) &=&\epsilon_k+
  \frac{p-p_k}{c^2} \nonumber\\
  &&+ \frac{p_k\Upsilon_k}{(1+\upsilon_{k+1})c^2}
  \left[\left(\frac{p}{p_k}\right)^{1+\upsilon_{k+1}}-1\right]\quad
\end{eqnarray}
in the pressure subdomain $p_k\leq p < p_{k+1}$.  The constants
$\Upsilon_k$ and $\epsilon_k$ are determined iteratively by the
recursion relations
\begin{eqnarray}
  \Upsilon_k &=& \Upsilon_{k-1}\left(\frac{p_k}{p_{k-1}}\right)^{\upsilon_{k}},
    \\
\epsilon_k &=&\epsilon_{k-1} + \frac{p_k-p_{k-1}}{c^2}\nonumber\\
&&+ \frac{p_{k-1}\Upsilon_{k-1}}{(1+\upsilon_{k})c^2}
  \left[\left(\frac{p_k}{p_{k-1}}\right)^{1+\upsilon_{k}}-1\right],
\end{eqnarray}
with $\Upsilon(p_0)=e^{\upsilon_0}$, which enforce continuity at the
pressure subdomain boundaries.  The remaining constants $\upsilon_k$
are the independent parameters that determine the equation of state in
each pressure subdomain.

Spectral representations of $\Upsilon(p)$ can also be
constructed, for example by setting
\begin{eqnarray}
  \Upsilon(p)=\exp\left[\sum_k \upsilon_k \Phi_k(p)\right].
  \label{e:UpsilonP}
\end{eqnarray}
The graphs in Fig.~\ref{f:SoundSpeeds} suggest that a reasonable
choice for $\Phi_k(p)$ is $\Phi_k(p)=[\log(p/p_0)]^k$, the same as
those used to represent $\Gamma(p)$ in Sec.~\ref{s:FaithfulEOS}.  The
associated representation of $\epsilon(p)$ is obtained from
Eq.~(\ref{e:EpsilonVelocity}):
\begin{eqnarray}
&&  \epsilon(p,\upsilon_k) = \frac{1}{c^2}+\frac{p-p_0}{c^2}
  \nonumber\\
  &&\qquad\quad+\frac{1}{c^{2}}
  \int_{p_0}^p\exp\left\{{\sum_k\upsilon_k\left[\log\left(
        \frac{p'}{p_0}\right)\right]^k}\right\}\,dp'.\qquad
\end{eqnarray}
Unfortunately, the integral in this expression can not be done
analyticly.  However the integrand is very smooth, and numerical
integration (e.g. using Gaussian quadrature) can be done very
accurately and efficiently.

The detailed discussion of how accurately and efficiently the causal
piecewise-analytical (PB-C-PA) and the causal spectral (PB-C-S)
representations\footnote{The notation PB-C-PA is used to denote the
  pressure based causal piecewise-analytical representation, while
  PB-C-S denotes the pressure-based causal spectral representation.}
work for these various equations of state is delayed to
Section~\ref{s:RealisticEOS}.  However to summarize, those tests
reveal that the PB-C-PA representations are much more efficient than
the non-causal PB-PA representations, while the PB-C-S representations
are roughly comparable to the PB-S representations.  Thus,
representations that enforce causality can be used without sacrificing
accuracy or efficiency.


\section{Enthalpy-Based Representations}
\label{s:EnthalpyBasedForms}

The causal representations described in Sec.~\ref{s:CausalEOS} are
based on the pressure-based forms of the equation of state,
$\epsilon=\epsilon(p)$.  These are quite useful for many purposes,
however, for some applications this standard form is not ideal.  For
example, a very useful form of the relativistic stellar structure
equations requires the equation of state to be expressed in terms of
the relativistic enthalpy, $h$~\cite{Lindblom1992}.  For applications
such as this, $\epsilon=\epsilon(p)$ must be re-written as a pair of
equations $\epsilon=\epsilon(h)$ and $p=p(h)$, where $h$ is defined as
\begin{eqnarray}
h(p) = \int_0^p \frac{dp'}{\epsilon(p')c^2+ p'}.
\label{e:EnthalpyDef}
\end{eqnarray} 
The needed expressions, $\epsilon=\epsilon(h)$ and $p=p(h)$, are
constructed by inverting $h=h(p)$ from Eq.~(\ref{e:EnthalpyDef}) to
obtain $p=p(h)$, and then composing the result with, $\epsilon(p)$, to
obtain $\epsilon(h)=\epsilon[p(h)]$.  

The transformations needed to construct $\epsilon=\epsilon(h)$ and
$p=p(h)$ from $\epsilon=\epsilon(p)$ are difficult to perform
numerically in an efficient and accurate way.  Therefore directly
constructing parametric enthalpy-based representations is preferable.
This can be done using methods analogous to those described in
Secs.~\ref{s:FaithfulEOS} and~\ref{s:CausalEOS}.  Enthalpy-based
representations have been reported in Ref.~\cite{Lindblom2010} that
satisfy the thermodynamic stability condition but not the causality
condition.  Those derivations will not be repeated here.  Instead, the
focus here will be on the construction of enthalpy-based causal
representations.

The functions $p(h)$ and $\epsilon(h)$ from the enthalpy-based form of
the equation of state satisfy the system of ordinary differential
equations,
\begin{eqnarray}
\frac{d\epsilon}{dh} &=& \frac{\epsilon\,c^2 + p}{c^2}
\left[1+\Upsilon(h)\right],
\label{e:epseq}\\
\frac{dp}{dh} &=& \epsilon\, c^2+ p,
  \label{e:peq}
\end{eqnarray}
which follow from the definitions of $\Upsilon$ in
Eq.~(\ref{e:EpsilonP}), and $h$ in Eq.~(\ref{e:EnthalpyDef}).  These
equations imply that there is a one to one correspondence between
equations of state, $p=p(h)$ and $\epsilon=\epsilon(h)$, and the
velocity function $\Upsilon(h)$.  Every equation of state determines
$\Upsilon(h)$; and conversely, every function $\Upsilon(h)$ determines
an equation of state through Eqs.~(\ref{e:epseq}) and (\ref{e:peq}).
Causal representations of the equation of state can therefore be
constructed from representations of non-negative functions
$\Upsilon(h)$.  These representations can be constructed using either
the piecewise-analytical or the spectral approaches.

\begin{figure}[!ht]
\centerline{\includegraphics[width=3in]{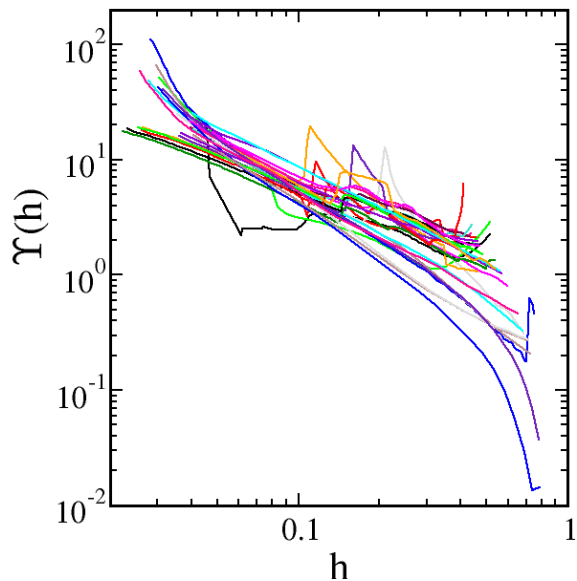}}
\caption{\label{f:SoundSpeedsh} Curves illustrate the sound speed
  function $\Upsilon(h)$ for 27 nuclear-theory model equations of
  state that satisfy the causality condition, $\Upsilon(h)\geq 0$, in
  the enthalpy domain of interest to neutron-star interiors.}
\end{figure}

Piecewise-analytical enthalpy-based representations can be constructed
using the same approach as the pressure-based representations given in
Secs.~\ref{s:FaithfulEOS} and \ref{s:CausalEOS}.  The first step is to
divide the domain of enthalpies relevant for the high density portion
of a neutron-star core, $[h_0,h_\mathrm{max}]$, into $n$ subdomains
with $h_0 < h_1 < ... < h_{n-1} < h_\mathrm{max}$.  The second step is
to choose analytical functions $\Upsilon(h,\upsilon_k)$ to approximate
$\Upsilon(h)$ in each subdomain.  The challenge is to find analytical
functions that are reasonably good approximations in each subdomain,
and that are simple enough to allow Eqs.~(\ref{e:epseq}) and
(\ref{e:peq}) to be solved analyticly for $\epsilon(h,\upsilon_k)$ and
$p(h,\upsilon_k)$.  Figure~\ref{f:SoundSpeedsh} shows $\Upsilon(h)$ in
the relevant range of $h$ for 27 causal nuclear-theory model equations
of state.  These graphs show that $\log\Upsilon$ is more or less
proportional to $-\log h$ for these equations of state.  This fact,
together with the need to have simple analytical functions having
analytical integrals, leads to the following choice for
$\Upsilon(h,\upsilon_k)$,
\begin{eqnarray}
  \Upsilon(h,\upsilon_k)=\frac{\upsilon_k+2(h_{k+1}-h)}{h},
  \label{e:UpsilonhDef}
\end{eqnarray}
in the subdomain $h_k\leq h < h_{k+1}$.  These functions are
non-negative within each subdomain so long as non-negative
adjustable parameters are chosen, $\upsilon_k\geq 0$.

The piecewise-analytical representation of the equation of state,
$\epsilon(h,\upsilon_k)$ and $p(h,\upsilon_k)$, that corresponds
to the $\Upsilon(h,\upsilon_k)$ given in Eq.~(\ref{e:UpsilonhDef})
is determined by solving Eqs.~(\ref{e:epseq}) and (\ref{e:peq}).
The general solution to these equations can be reduced to quadrature:
\begin{eqnarray}
p(h)&=&p_0+ \left(\epsilon_0\,c^2+p_0\right)   \int_{h_0}^h\mu(h')\,dh',
\label{e:PressueH}\\
\epsilon(h)& =& -p(h)\,c^{-2}+\left(\epsilon_0+p_0\,c^{-2}\right)\mu(h),
\label{e:EnergyH}
\end{eqnarray}
where $\mu(h)$ is defined as, 
\begin{eqnarray}
\mu(h) &=& \exp\left\{ \int_{h_0}^h \left[2+\Upsilon(h')\right]dh'\right\}.
\label{e:MuHDef}
\end{eqnarray}
The constants $p_0$ and $\epsilon_0$ are defined by $p_0=p(h_0)\geq 0$
and $\epsilon_0=\epsilon(h_0)\geq 0$ respectively.  Inserting the
expression for $\Upsilon(h,\upsilon_k)$ from Eq.~(\ref{e:UpsilonhDef})
into these integrals gives the following expressions for the
equation of state,
\begin{eqnarray}
  p(h,\upsilon_k)&=&p_k+\frac{\left(\epsilon_kc^2+p_k\right)h_k}
  {\upsilon_k+2h_{k+1}+1}
  \left[\left(\frac{h}{h_k}\right)^{\upsilon_k+2h_{k+1}+1}\!\!\!\!\!\!
    -1\right],\nonumber\\\\
  \epsilon(h,\upsilon_k)&=&-p(h,\upsilon_k)c^{-2}
  +\left(\epsilon_k+p_k c^{-2}\right)
  \left(\frac{h}{h_k}\right)^{\upsilon_k+2h_{k+1}}\!\!\!,\nonumber\\
\end{eqnarray}
in the subdomain $h_k\leq h < h_{k+1}$.  The constants
$p_k=p(h_k,\upsilon_k)$ and $\epsilon_k=\epsilon(h_k,\upsilon_k)$ are
determined from the recursion relations,
\begin{eqnarray}
  p_{k+1}&=&p_k+\frac{\left(\epsilon_kc^2+p_k\right)h_k}{\upsilon_k+2h_{k+1}+1}
  \left[\left(\frac{h_{k+1}}{h_k}\right)^{\upsilon_k+2h_{k+1}+1}\!\!\!\!
    -1\right],\nonumber\\\\
  \epsilon_{k+1}&=&-p_{k+1}c^{-2}
  +\left(\epsilon_k+p_k c^{-2}\right)
  \left(\frac{h_{k+1}}{h_k}\right)^{\upsilon_k+2h_{k+1}}.\nonumber\\
\end{eqnarray}

Causal enthalpy-based representations of the equation of state
can also be constructed using spectral methods.  The basic
idea is to represent the velocity function $\Upsilon(h)$ as 
a spectral expansion:
\begin{equation}
  \Upsilon(h,\upsilon_k)=\exp\left[\sum_k \upsilon_k\,\Phi_k(h)\right],
    \label{e:UpsilonSpectralH}
\end{equation}
where the $\Phi_k(h)$ is any complete set of basis functions on the
domain $\left[h_0,h_\mathrm{max}\right]$.  Figure~\ref{f:SoundSpeedsh}
shows that the $\Upsilon(h)$ for the collection of 27 causal
nuclear-theory model equations of state are fairly simple functions of
$\log h$.  Therefore, following the approach used in
Ref.~\cite{Lindblom2010} we take the spectral basis functions
$\Phi_k(h)$ to be $\Phi_k(h)=[\log(h/h_0)]^k$. The expression for
$\Upsilon(h,\upsilon_k)$ then becomes:
\begin{eqnarray}
  \Upsilon(h,\upsilon_k)
  = \exp\left\{\sum_k \upsilon_k \,\left[\log\left(\frac{h}{h_0}
\right)\right]^k\right\}.
\label{e:UpsilonH}
\end{eqnarray}
While the quadratures in Eqs.~(\ref{e:PressueH}) and (\ref{e:MuHDef}) can
not be done analyticly for the spectral expansion defined in
Eqs.~(\ref{e:UpsilonH}), they can be done numerically very efficiently
and accurately using Gaussian quadrature.

The detailed discussion of how accurately and efficiently the
enthalpy-based causal piecewise-analytical (HB-C-PA) and the causal
spectral (HB-C-S) representations\footnote{The notation HB-C-PA is
  used to denote the enthalpy-based causal piecewise-analytical
  representation, while HB-C-S denotes the enthalpy-based causal
  spectral representation.}  work for these various equations of state
is delayed to Section~\ref{s:RealisticEOS}.  However to summarize,
those tests reveal that the HB-C-PA representations are roughly
comparable to the non-causal PB-PA representations (somewhat better
for $n\geq 4$, somewhat worse for $n\leq 3$), while the HB-C-S
representations are roughly comparable to the HB-S representation.
Thus enthalpy-based representations that enforce causality can
be used without sacrificing accuracy or efficiency.

 
\section{Testing the Representations}
\label{s:RealisticEOS}

This section presents numerical tests of the accuracy and efficiency
of the various parametric representations of neutron-star equations of
state described in
Secs.~\ref{s:FaithfulEOS}--\ref{s:EnthalpyBasedForms}.
Representations of 27 nuclear-theory model neutron-star equation of
state models are constructed and then compared with the originals to
measure their accuracy and efficiency.  The 27 model equations of
state used in these tests are the causality satisfying subset of the
collection of 34 nuclear-theory based models used by Read, et
al.~\cite{Read:2008iy} in their study of the piecewise-polytropic
representations.  The particular equations of state excluded from their
collection are APR2, APR3, APR4, WFF1, WFF2 ENG, and ALF1, each of
which violates the causality condition $dp/d\epsilon\leq c^2$ for some
range of densities.

\begin{figure}[t]
\centerline{\includegraphics[width=3in]{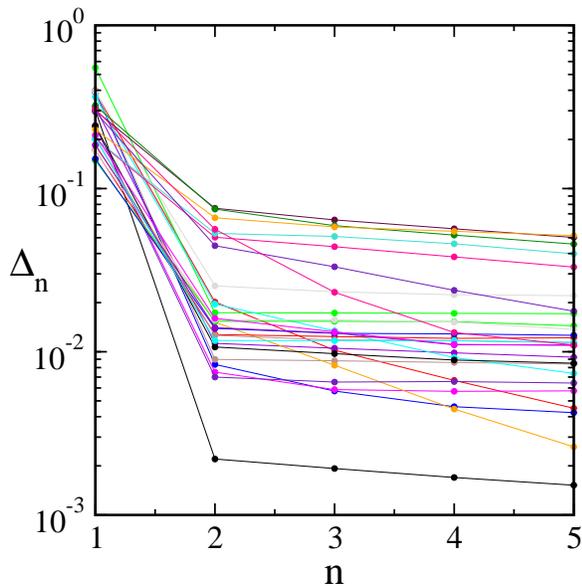}}
\caption{\label{f:p_causal_simp_chisqr_nlambda} Residuals $\Delta_n$
  are illustrated as functions of $n$ (the number of parameters
  $\upsilon_k$ used) for the pressure-based causal piecewise-analytical
  (PB-C-PA) representations of 27 causal nuclear-theory model
  equations of state.}
\end{figure}

\begin{figure}[t]
\centerline{\includegraphics[width=3in]{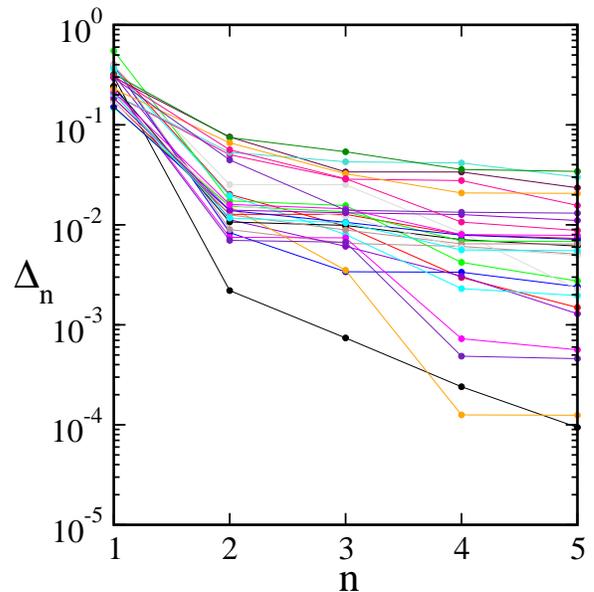}}
\caption{\label{f:p_causal_spec_chisqr_nlambda} Residuals $\Delta_n$
  are illustrated as functions of $n$ (the number of parameters
  $\upsilon_k$ used) for the pressure-based causal spectral (PB-C-S)
  representations of 27 causal nuclear-theory model equations of
  state.}
\end{figure}

Each of the nuclear-theory model equations of state consists of a
table of energy-density pressure pairs: $\{\epsilon_i,p_i\}$ for
$1\leq i \leq N$.  A parametric representation of one of these
equations of state consists of a function $\epsilon(p,\upsilon_k)$
that gives the energy density as a function of the pressure, and also
the parameters $\upsilon_k$ for $1\leq k \leq n$.  The accuracy of a
particular representation is determined by evaluating the
dimensionless error residual $\Delta_n(\upsilon_k)$, defined by
\begin{eqnarray}
\Delta^2_n(\upsilon_k)=\sum_{i=1}^N\frac{1}{N}\left\{
\log\left[\frac{\epsilon(p_i,\upsilon_k)}
  {\epsilon_i}\right]\right\}^2.
\label{e:ResidualDef}
\end{eqnarray}
The optimal choice of the parameters $\upsilon_k$ to represent a
particular equation of state $\{\epsilon_i,p_i\}$ is obtained by
minimizing $\Delta_n(\upsilon_k)$ with respect to variations in each
of the parameters $\upsilon_k$.  This minimization is carried out
numerically for these tests using the Levenberg-Marquardt
method~\cite{numrec_f}.  The minimum values of $\Delta_n$ obtained in
this way are illustrated for the pressure-based causal
piecewise-analytical (PB-C-PA) representation in
Fig.~\ref{f:p_causal_simp_chisqr_nlambda} and the pressure-based
causal spectral (PB-C-S) representation in
Fig.~\ref{f:p_causal_spec_chisqr_nlambda}.  Each of the 27 curves in
these figures represents the residual $\Delta_n$ as a function of the
number of parameters $n$ for one of the causal nuclear-theory model
equations of state.  These curves generally show convergence,
i.e. $\Delta_{n}\geq \Delta_{n+1}$, thus showing that the minimization
methods used in these tests are effective.  As expected, the different
curves converge at different rates for the various equations of state,
depending on how complicated the structure of the function
$\Upsilon(p)$ is for each case.  Also as expected, the average
residuals $\Delta_n$ for the spectral representations in
Fig.~\ref{f:p_causal_spec_chisqr_nlambda} are smaller (for given value
of $n$) than their counterparts for the piecewise-analytical
representations in Fig.~\ref{f:p_causal_simp_chisqr_nlambda}.

\begin{figure}[t]
\centerline{\includegraphics[width=3in]{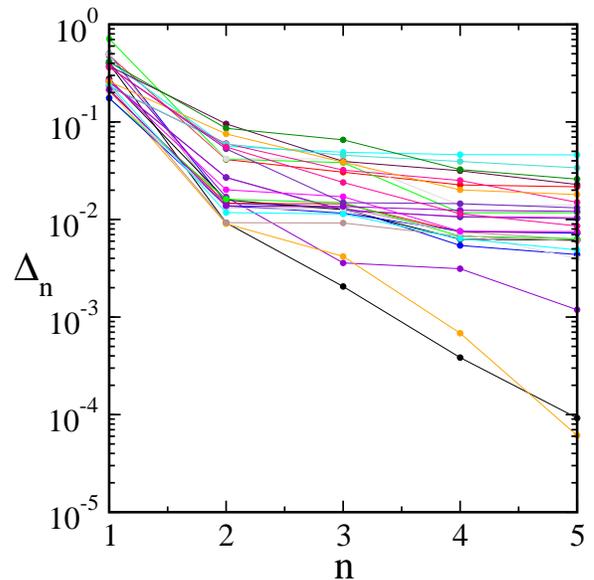}}
\caption{\label{f:h_causal_spec_chisqr_nlambda} Residuals $\Delta_n$
  are illustrated as functions of $n$ (the number of parameters
  $\upsilon_k$ used) for the enthalpy-based causal spectral (HB-C-S)
  representations of 27 causal nuclear-theory model equations of
  state.}
\end{figure}

Figure~\ref{f:h_causal_spec_chisqr_nlambda} illustrates the residuals
$\Delta_n$ for the enthalpy-based causal spectral (HB-C-S)
representation discussed in Sec.~\ref{s:EnthalpyBasedForms}.  The
residuals in this case are computed in much the same way as those for
the pressure-based representations.  The only difference is the
$\epsilon(h_i,\upsilon_k)$ that appears in error residual
\begin{eqnarray}
\Delta^2_n(\upsilon_k)=\sum_{i=1}^N\frac{1}{N}\left\{
\log\left[\frac{\epsilon(h_i,\upsilon_k)}
  {\epsilon_i}\right]\right\}^2
\label{e:ResidualDef}
\end{eqnarray}
now depends on $h_i$, the enthalpy corresponding to the point
$\{\epsilon_i,p_i\}$ for the particular equation of state.  Comparing
Figs.~\ref{f:p_causal_spec_chisqr_nlambda} and
\ref{f:h_causal_spec_chisqr_nlambda} shows that the accuracy and
efficiency of the enthalpy-based spectral representations are
comparable to their pressure-based counterparts.


\begin{figure}[t]
\centerline{\includegraphics[width=3in]{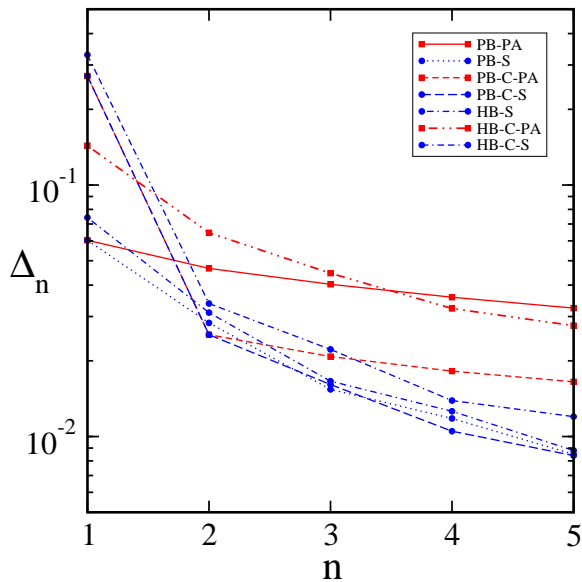}}
\caption{\label{f:x2_average} Convergence of the average residuals
  $\Delta_n$ is illustrated for the seven different representations
  studied here.  Piecewise-analytical representations are shown
  with square data points (and colored red), while the spectral representations
are shown with round data points (and colored blue).}
\end{figure}

Representations of the 27 causal nuclear-theory model equations of
state were constructed in this way for each of the seven different
methods described in
Secs.~\ref{s:FaithfulEOS}--\ref{s:EnthalpyBasedForms}.  In particular
representations were constructed for the pressure-based
piecewise-analytical (PB-PA) and the pressure-based spectral (PB-S)
representations described in Sec.~\ref{s:FaithfulEOS}, the
pressure-based causal piecewise-analytical (PB-C-PA) and the
pressure-based causal spectral (PB-C-S) representations described in
Sec.~\ref{s:CausalEOS}, and finally for the enthalpy-based spectral
(HB-S), the enthalpy-based causal piecewise-analytical (HB-C-PA), and
the enthalpy-based causal spectral (HB-C-S) representations discussed
in Sec.~\ref{s:EnthalpyBasedForms}.  Some of these representations
(PB-PA, PB-S, and HB-S) had been discussed previously in the
literature (see e.g. Ref.~\cite{Lindblom2010}) so detailed convergence
plots are not given here for those cases.  Instead the average values
of $\Delta_n$ (where the averages are taken over the 27 causal
equation of state models) are illustrated in Fig.~\ref{f:x2_average}.
This figure shows clearly that the spectral representations (with
circular data points and colored blue) are more efficient on average
than the piecewise-analytical representations (with square data points
and colored red).  This figure also shows that the pressure-based
causal piecewise-analytical representation (PB-C-PA) is much more
efficient on average than either of the other two piecewise-analytical
representations (PB-PA and HB-C-PA) considered here.  And, this figure
reveals that all the spectral representations are more or less
comparable, with the pressure-based causal spectral (PB-C-S)
representation being slightly more efficient than the others studied
here.
%

\section{Discussion}
\label{s:Discussion}

The graphs in
Figs.~\ref{f:p_causal_simp_chisqr_nlambda}--\ref{f:h_causal_spec_chisqr_nlambda}
illustrate the convergence of the error residuals $\Delta_n$ as the
number of parameters $n$ is increased for three new causal
representations of neutron-star equations of state presented here.
Each curve in each figure represents one of the 27 causal
nuclear-physics model equations of state included in this study.
These graphs show that there is a fairly large range of convergence
rates for these representations among the different equation of state
models.  The differences in convergence rates arise from the
differences in the equation of state functions $\Upsilon(p)$ or
$\Upsilon(h)$ illustrated in Figs.~\ref{f:SoundSpeeds} and
\ref{f:SoundSpeedsh}.  Those equations of state with $\Upsilon$ having
more structure, e.g. including discontinuities from phase transitions,
have representations that converge more slowly than smooth equations
of states having less structure.  These graphs also show that even the
most difficult cases studied here (i.e. those with strong phase
transitions) have errors $\Delta_n< 0.1$, for $n\geq 2$, and
$\Delta_5\leq 0.05$.  These graphs also show that the spectral
representations converge more rapidly than the piecewise analytical
representations, even for the more difficult cases with phase
transitions.  The relationship between the equation of state structure
and the efficiency of representations was studied in some detail in
Ref.~\cite{Lindblom2010}.  That earlier work focused on non-causal
representations, however, the relationship between the structure of
the equation of state and the convergence rate of these types of
representations is the same in the cases studied here.

The most widely used representation of the neutron-star equation of
state (at the present time) is the piecewise-polytropic
representation~\cite{Vuille1995, VuilleIpser1999, Read:2008iy}.  The
most carefully studied version of this representation uses four
adjustable parameters~\cite{Read:2008iy}, whose best-fit
representation has an error residual $\Delta_4=0.0383$ when averaged
over 34 nuclear-theory model equations of state~\cite{Lindblom2010}.
While this representation is essentially equivalent to the
pressure-based piecewise-analytical representation (PB-PA) discussed
in Sec.~\ref{s:FaithfulEOS}, there are some minor differences.  The
solid curve in Fig.~\ref{f:x2_average} shows the convergence of the
average residuals $\Delta_n$ for the PB-PA representation, which has
the value $\Delta_4=0.0358$ for the four-parameter case.  These
results differ slightly from the earlier work for two reasons.  First,
the collection of 27 causal nuclear-theory model equations of state
used in the tests here is a subset of the 34 equations of state used
in the earlier studies.  And second, the four parameters adjusted in
the earlier study to obtain the best-fit representation are the three
subdomain adiabatic index parameters, $\gamma_1$, $\gamma_2$ and
$\gamma_3$, plus the location of the first pressure subdomain boundary
$p_1$.  In the present work the locations of the subdomain boundaries
are uniformly spaced and fixed, while the $n$ adiabatic index
parameters $\gamma_k$ for $k=1,...,n$ are adjusted to obtain the best
fits. 

One notable feature of the results of the convergence tests performed
here is how slowly $\Delta_n$ decreases for the PB-PA representation
as the number of parameters is increased.  Figure~\ref{f:x2_average}
shows that while the residual $\Delta_1$ for the one parameter version
of PB-PA is actually the smallest, its residuals $\Delta_n$ become the
largest when $n\geq 4$.  Almost all of the other representations
(except HB-C-PA) have smaller two parameter residuals $\Delta_2$ than
$\Delta_4$ for the widely used four-parameter version of PB-PA.  The
PB-PA representation is therefore (along with HB-C-PA) the least
accurate and efficient of the representations studied here.  These
representations therefore fail to satisfy the basic Principle~\#3,
efficiency, from the discussion in Sec.~\ref{s:introduction}.  The
PB-C-PA representation is much more efficient than PB-PA, and has the
added advantage of automatically enforcing causality.  Therefore if a
piecewise-analytical representation is desired (e.g. because it is
easier to code up, or because it is more efficient to use numerically
than the spectral representations) the PB-C-PA representation
described in Sec.~\ref{s:CausalEOS} is probably a better choice than
PB-PA.  The most efficient representations, however, are the spectral
representations, PB-S, PB-C-S, HB-S and HB-C-S.  The most efficient of
these is the pressure-based causal spectral representation PB-C-S.
This representation also satisfies Principles~\#1 and \#2 of the
general discussion in Sec.~\ref{s:introduction}.  This study therefore
suggests that the PB-C-S representation is probably the best choice
for a representation of the neutron-star equation of state available
at this time.


\acknowledgments

I thank Steve Lewis and Fridolin Weber for helpful comments and
suggestions on an early draft of this paper.  This research was
supported in part by grants PHY-1604244 and DMS-1620366 from the
National Science Foundation.

\vfill\eject
\bibstyle{prd} 
\bibliography{../References/References}
\end{document}